\documentclass{iau}

\usepackage{graphicx}
\usepackage{amsmath}
\usepackage{multirow}
\usepackage{aas_macros}
\usepackage{amsfonts}
\usepackage[utf8]{inputenc}
\usepackage[T1]{fontenc}
\usepackage[english]{babel}
\usepackage{bm}
\usepackage{empheq}
\usepackage{xcolor}
\usepackage{soul}

\DeclareMathAlphabet{\mathcal}{OMS}{cmsy}{m}{n}

\newcommand\microsec{\textnormal{~$\mu$s}}
\newcommand{\vect}[1]{\bm{#1}}
\def\dd{\mathrm{d}}

\def\A{\mathrm{e}}
\def\B{\mathrm{l}}  
\def\pptimeA{\tau_\A}
\def\pptimeB{\tau_\B}
\def\tcb{t}

\def\tcg{T}
\def\tcl{\mathcal{T}}
\def\A{\mathrm{e}}
\def\B{\mathrm{l}}  
\def\posb{x}
\def\posg{X}  
\def\posl{\mathcal{X}}  
\def\xl{\mathbf{\posl}}
\def\xg{\vect{\posg}}
\def\xb{\vect{\posb}}

\def\xgA{\vect{\posg}_\A}

\def\xlB{\vect{\posl}_\B}

\def\vitb{v}
\def\vitg{V}
\def\vitl{\mathcal{V}}

\def\vb{\vect{\vitb}}

\def\vgA{\vect{\vitg}_\A}

\def\vitlB{\vitl_\B}

\def\terre{\mathrm{E}}
\def\lune{\mathrm{L}}

\def\poteff{\mathsf{W}}
\def\potGCRSeff{W}
\def\potM{\poteff_{\mathrm{\lune}}}

\def\potGCRSE{\potGCRSeff_{\mathrm{E}}}

\def\pottide{\poteff_{\mathrm{tidal}}}
\def\potGCRStide{\potGCRSeff_{\mathrm{tidal}}}

\def\potiner{\poteff_{\mathrm{iner}}}
\def\potGCRSiner{\potGCRSeff_{\mathrm{iner}}}

\def\TCL{\mathrm{TCL}}
\def\TCG{\mathrm{TCG}}
\def\TCB{\mathrm{TCB}}
\def\TL{\mathrm{TL}}
\def\TT{\mathrm{TT}}

\def\TDB{\mathrm{TDB}}

\begin{document}




\title{Lunar Time}

\author{P.~Defraigne$^1$,   F.~Meynadier$^3$, A.~Bourgoin$^2$}

\affiliation{$^1$ Royal Observatory of Belgium}
\affiliation{$^2$ Time Department, BIPM}
\affiliation{$^3$ LTE, Observatoire de Paris}

\begin{abstract}
The regain of interest in Moon exploration has substantially grown in the last years. For this reason, the space agencies consider the development of 
a precise navigation and positioning service similar to the Earth GNSS. Aiming at some meter accuracy, this requires to set up a relativistic lunar reference frame, with an associated coordinate time. If the IAU already defined the Lunar Coordinate Time TCL, there is still some freedom in the choice of the coordinate timescale to be adopted as reference on or around the Moon.
This paper proposes a trade-off analysis of different possible options for this reference time scale.  It shows that  TCL  is the best option to be used as practical time reference on the Moon, without the need to define a new time scale based on a scaling of TCL.

\end{abstract}


\maketitle

\section{Introduction}

In view of the future institutional and private lunar exploration missions,
space agencies are developing lunar communication and navigation services that would
allow positioning accuracy on the Moon as well as during the final descent of landers on the  surface. In
particular, NASA is developing the LunaNET \citep{Israel2020}, and ESA is developing
the Moonlight program \citep{Moonlight2022}, both aiming to provide communication
and Position Navigation and Timing (PNT) services to future
lunar missions, first for robotic landers and then for manned missions. China also proposes its own “Queqiao comprehensive constellation
for communications, navigation and remote sensing” \citep{China2019}.
In order to allow the users to increase their PNT accuracy thanks to combining the 
signals from different satellites, it is mandatory that all the different satellites 
use the same coordinate system and reference time scale.  This stresses the need to define such systems. 
Given that the objective is to offer precise positioning to within a few meters, time measurement must be accurate to within a few nanoseconds, and  including general relativity cannot be avoided. 

The IAU  resolution II \citep{IAUGA2024} adopted during the 2024 General Assembly recommends constructing a Lunar Celestial Reference System (LCRS), with its
coordinate time designated Lunar Coordinate Time (TCL), using the same technique
as to construct the GCRS and TCG for the Earth, and keeping the unit of measurement of TCL
consistent with the SI second. The IAU resolution III \citep{IAUGA2024}, also noting that increased robotic and human activity on the Moon in the near future requires a
practical and internationally recognized lunar reference time scale, and that is suitable as the
basis for scientific measurements there, recommends that
the relationships between the possible versions of a lunar reference time scale and other
time scales, in particular a lunar coordinate time and UTC, are pursued in collaborative
agreement among the relevant international organizations.

This paper presents a trade-off analysis of different possible options
for a reference lunar time.  It also proposes how to provide to the lunar users
a predictable difference to the Earth time reference, i.e. UTC,  at least up to a certain level of accuracy, 
  in order to allow the lunar explorers  to coordinate their
activities with Earth-based control centers. The discussion proposed here relies on  the theoretical developments  detailed in \cite{Bourgoin2025}.

\section{Coordinate time, proper time}

\subsection{Coordinate times TCB, TCG, TCL}

For an observer in the Solar System, the IAU Resolution B1.3 (2000) \citep{IAUGA2000} defined
the global Barycentric Celestial Reference System BCRS, with its origin at the Solar System barycenter, and its associated
coordinate time TCB. However, for an observer on or around the Earth, it is more
appropriate to rely on a local coordinate system whose origin is at the center-of-mass
of the Earth. Accordingly, the IAU Resolution B1.3 (2000) \citep{IAUGA2000} also defined
the Geocentric Celestial Reference System GCRS with its associated coordinate time TCG. The GCRS is a non-rotating
reference system, and can be considered as quasi-inertial in the vicinity of the Earth, where the external potential reduces essentially to tidal terms. It
is classically used in the region of space from the Earth up to the geostationary altitude.
The IAU 2024 resolution II \citep{IAUGA2024} recommends constructing a Luni-centric Celestial Reference System LCRS with TCL
as the coordinate time, using the same approach as for constructing the GCRS,
with TCL as the coordinate time. It also recommends that the unit of measurement of
TCL be consistent with the SI second.
The relations between the different time scales are  provided by the IERS Conventions \citep{2010ITN....36....1P}:
\begin{equation}
[\tcg -\tcb]_{(\tcb,\xb)} = - \frac{1}{c^{2}} \left\{\int_{t_0}^{t}\left[\frac{\vitb^2_\terre(\tcb')}{2}+w_{\terre}(\xb_\terre(\tcb'))\right] \dd \tcb' + \vb_\terre (\tcb) \cdot [\xb-\xb_\terre(\tcb)] \right\}+\mathcal{O}(c^{-4}) \, ,
  \label{eq:TCB-TCG}
\end{equation}
and
\begin{equation}
[\tcl -\tcb]_{(\tcb,\xb)} = -\frac{1}{c^2} \left\{\int_{t_0}^{\tcb}\left[\frac{\vitb^2_\lune(\tcb')}{2}+w_{\lune}(\xb_\lune(\tcb'))\right] \dd \tcb' + \vb_\lune (\tcb) \cdot [\xb-\xb_\lune(\tcb)] \right\}+\mathcal{O}(c^{-4}) \, ,
  \label{eq:TCB-TCL}
\end{equation}
where $c$ is the speed of light, $\tcb$ is TCB, $\tcg$ is TCG, $\tcl$ is TCL, $\xb$ and $\vb$ are the position and velocity in the BRCS, $\vitb=\vert \vb \vert$,  $\terre$ denotes the geocenter and $\lune$ the center-of-mass of the Moon,  $w_{\terre}$ is the Newtonian potential, evaluated at the geocenter, due to all bodies in the solar system besides Earth, and $w_{\lune}$ is the Newtonian potential, evaluated at the center of mass of the Moon, due to all bodies in the solar system except the Moon. The reference epoch $t_0$ is 1977 January 1st, $0^{\mathrm{h}}0^{\mathrm{m}}32.184^{\mathrm{s}}$ in TCB, or 1977 January 1st, $0^{\mathrm{h}}0^{\mathrm{m}}0^{\mathrm{s}}$ TAI.

\subsection{Proper time around the Earth}
A clock can usually be considered as representing an observer, meaning that  its spatial coordinates in a given reference system might be seen as functions of the corresponding coordinate time. Two clocks in a given reference system are synchronous if they have the same coordinate time. In the GCRS, the position of a clock `$\A$' is a given function of TCG, namely $\xgA=\xgA(\tcg)$; its coordinate velocity is thus $\vgA=\vgA(\tcg)$ with $\vgA = \dd\xgA/\dd\tcg$.  The proper time $\pptimeA $ of that clock  can be expressed in the IAU recommended framework as
\begin{equation}
\label{eq:dtaudTCG}
\frac{\dd \pptimeA - \dd \tcg}{\dd \tcg}=-\frac{1}{2c^2}\left[\vitg_\A^2(\tcg)+2\potGCRSeff(\tcg,\xgA)\right] + \mathcal{O}(c^{-4}) \, ,
\end{equation}
where the first term is the centrifugal potential, with $\vitg_\A = \vert \vgA \vert$, and where the effective potential $\potGCRSeff$ is given by
\begin{equation}
  \potGCRSeff(\tcg,\xgA) = \potGCRSE(\tcg,\xgA) + \potGCRStide(\tcg,\xgA) + \potGCRSiner(\tcg,\xgA) \, .
\end{equation}
where $\potGCRSE$ is the Newtonian potential of the Earth, $\potGCRStide$ is the tidal contribution from the  Moon, the Sun, and the other planets, and $\potGCRSiner$ is the inertial potential caused by the non-geodesic acceleration of the Earth center-of-mass (namely the couplings between the Earth non-spherical part of the gravity potential to external gravity fields from other bodies). When limiting the proper time to $10^{-17}$ in relative frequency accuracy, only the Newtonian potential of the Earth should be kept in equation \ref{eq:dtaudTCG}. 

\subsection{Coordinate times TT and TDB}
The sum of the centrifugal and effective potentials gives the gravity potential at any position in the rigidly rotating frame of the Earth. All ideal clocks located at rest on a mean equipotential surface in the Earth vicinity, will tick at the same rate in the GCRS (up to small periodic terms of order $10^{-17}$ in frequency), namely they will nearly beat the same rate in TCG. Note that of course real clocks will have rate differences due to their own limited accuracies.

As humanity was working with clocks on the surface of the Earth since the origin, a new coordinate time was defined by the IAU \citep{IAUGA2000}, named  the Terrestrial Time (TT), which is a scaled version of the coordinate time TCG, so that the mean rate of TT agrees as much as possible with that of the proper time of an observer located at a conventional reference potential $W_0$ which is close to the potential at the mean see level: 
\begin{equation}
\label{eq:TCG-TT}
 \TT - \TCG = - L_G \times (\mathrm{JD}_{\TT} -T_0) \times 86\,400\ \mathrm{s} \, 
\end{equation}
with $L_G = 6.969\,290\,134 \times 10^{-10}$ and where $\mathrm{JD}_{\TT}$ is the TT Julian date and $T_0 = 2\,443\,144.500\,372\,5$.

Let us note that according to \citet{Klioner2008}, scaling the coordinate time is necessarily accompanied by the corresponding scaling of spatial coordinates: $\xg^* = (1-L_G)\xg$, and mass parameter of celestial bodies: $(Gm)^* = (1-L_G) \, (Gm)$, where $G$ is the gravitational constant.  The scaled coordinate times and spatial coordinates can be seen as defining
a new reference system with coordinates $(c\tcg^*, \xg^{*} )$ where $\tcg^*$ denotes TT. 

IAU Resolution B3 (2006) also defined an additional coordinate time named Barycentric Dynamical Time (TDB)  as a scaled version of TCB with a scaling factor chosen in order to have the same average rate as TT at the geocenter:
\begin{equation}
\TDB - \TCB = - L_B \times (\mathrm{JD}_{\TCB} - T_0) \times 86\,400\ \mathrm{s} + \TDB_0 \, ,
\label{eq:TDB-TCB}
\end{equation}
where $\mathrm{JD}_{\TCB}$ is the TCB Julian date; $T_0$ was given earlier in Eq. \eqref{eq:TCG-TT}. $L_B$ and $\TDB_0$ are two defining constants: $L_B = 1.550\,519\,768 \times 10^{-8}$ and $\TDB_0 = -6.55 \times 10^{-5}\ \mathrm{s}$, respectively. This scaling of the coordinate time is also accompanied by the corresponding scaling of spatial coordinates: $\xb^{**} = (1-L_B)\xb$, and mass parameter of celestial bodies: $(Gm)^{**} = (1-L_B) \, (Gm)$.

\subsection{Proper time around the Moon}
\label{ProperMoon}
Similarly to the Earth, we can express the rate of proper time $\pptimeB$ of clock `$\B$' located on or around the Moon with respect to TCL:
\begin{equation}
  \frac{\dd \pptimeB - \dd \tcl}{\dd \tcl} = -\frac{1}{2c^2} \left[ \vitlB^2 + 2\poteff(\tcl,\xlB) \right] 
  \label{eq:dpptimedtcl}
\end{equation}
where $\xlB$ represents the clock position in the LCRS, and $\vitlB$ is the norm of the LCRS velocity of clock `$\B$'. The term $\vitlB^2/2$ is called the centrifugal potential at clock `$\B$' and the effective potential reads
\begin{equation}
  \poteff(\tcl,\xl) = \potM(\tcl,\xl) + \pottide(\tcl,\xl) + \potiner(\tcl,\xl) \, ,
  \label{eq:Wscal}
\end{equation}
with $\potM$ the lunar Newtonian potential, $\pottide$ the tidal potential (mainly from Earth and Sun), and $\potiner$ the inertial potential caused by the non-geodesic acceleration of the Moon center-of-mass. As for the Earth, the  inertial potential is negligible. Contrarily to the Earth, where the tidal potential is only periodic, $\pottide(\tcl,\xl)$ contains a constant term, due to the Moon spin orbit resonance 1:1 which renders one component of the tides raised by the Earth on the Moon permanent. Once inserted into Eq. \eqref{eq:dpptimedtcl},  this term contributes at the level of $2.37 \times 10^{-16}$ and depends on the location of the clock on the lunar surface.

\begin{figure}[t]
  \begin{center}
    \includegraphics[scale=0.45]{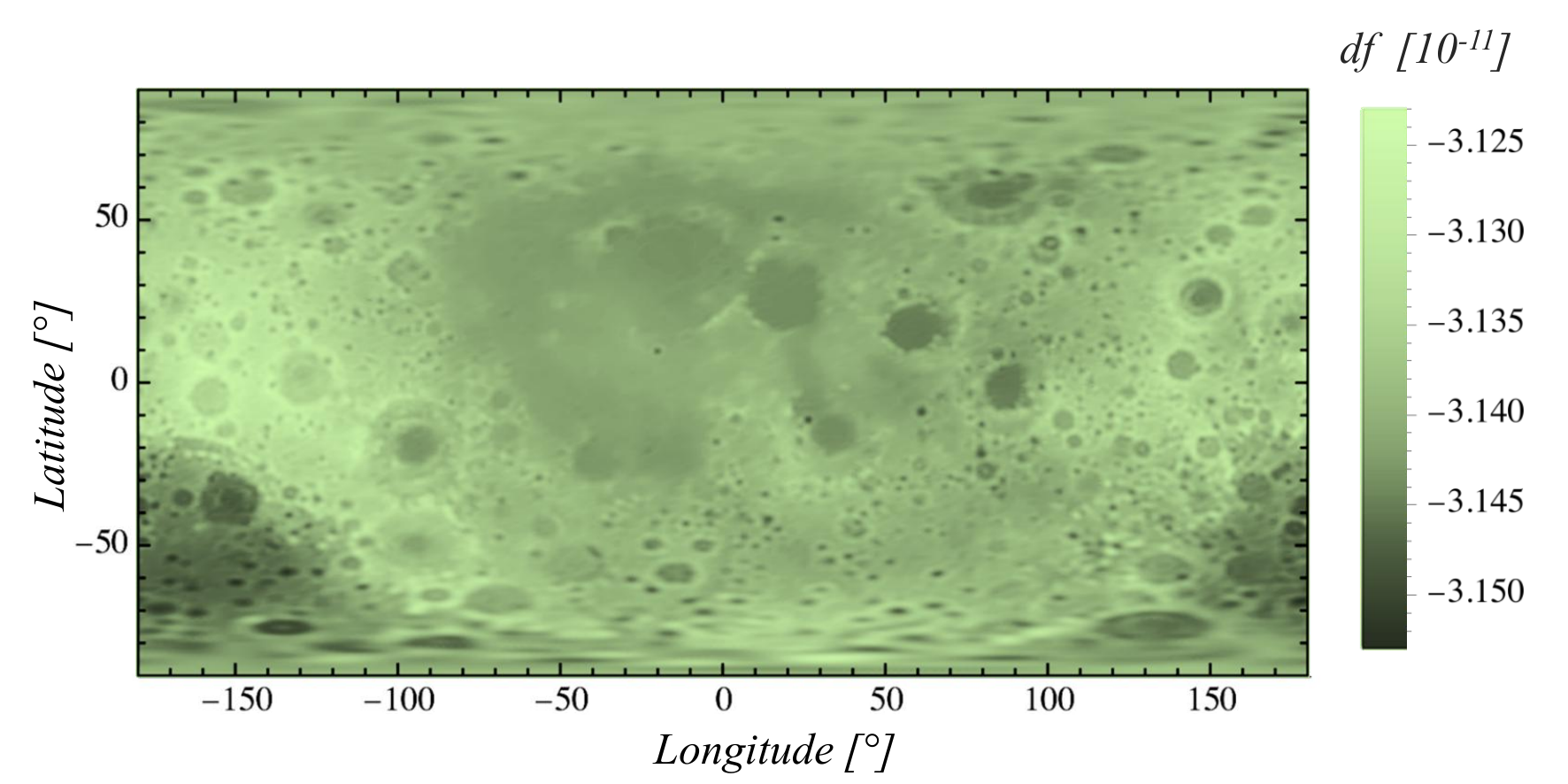}
  \end{center}
  \caption{Map of the relative frequency difference between the proper time of a clock at rest on the lunar surface and TCL. The origin of the longitude represents lunar prime meridian.}
  \label{fig:lunar_total}
\end{figure}

Considering the centrifugal potential and the gravitational potential from the Gravity Recovery and Interior Laboratory GRAIL \citep{Lemoine2014}, we can determine the  relative frequency difference between
proper time and TCL for a clock on the surface of the Moon. The clock altitude must furthermore be considered as the Newtonian potential depends also on the distance to the Moon center. Using the topography of the Moon obtained by the Lunar Orbiter Laser Altimeter LOLA \citep{2009ApOpt..48.3035R,2010SSRv..150..209S}, we can obtain relative frequency difference between the 
proper time and TCL,  depending on the position of the clock at rest on the Moon, as in  Figure \ref{fig:lunar_total}. 
The main effect  is at the level of $3.14 \times 10^{-11}$, it corresponds to the monopole $\poteff_{\mathrm{\lune}0}(\tcl,\xl)=G m_\lune / R_\lune$ where $m_\lune$ is the mass of the Moon, and $R_\lune$ is the mean Moon radius. The  clock relative frequency variations are within 3 $\times 10^{-13}$ along the surface, due to the high altitudes differences on the lunar topography.

\section{Difference between TCL and TT}
\label{differences}
Different authors recently provided theoretical differences between the rate of clocks on the Moon and on the Earth, using different approaches \citep{Kopeikin2024,Ashby2024,2025ApJ...985..140T}. As shown in \cite{Bourgoin2025}, these periodic terms depend on the coordinate system used --BCRS, GCRS, LCRS or even a coordinate system at the Earth-Moon barycenter as proposed by  \citet{Ashby2024}-- and are also specific to the locations of the clocks on    the Moon and  on the Earth,  as a consequence of equations \ref{eq:TCB-TCG} and \ref{eq:TCB-TCL}. Therefore, a universal difference between TCL and TT does not exist, and it is only possible to determine the difference between one clock realizing TCL on the Moon, and one clock realizing TT  on the Earth. This clock comparison must take into account the difference $\TCL-\TT$ which depends on the clock locations and on  the intermediate reference system which is used for the light travel time computation between the Earth station and the Moon station.  

On the Earth, TT  is realized by the Coordinated Universal Time (UTC, in which leap seconds are furthermore  inserted to maintain UTC close to the timescale associated with the Earth rotation angle UT1).  In view of the future communication between the Earth and the Moon, it is  needed to establish the relation between the local coordinate time employed around the Moon and UTC.    
 We expect  that UTC will
 remain the common operational timescale for humans on the Moon, as the
contact and coordination with Earth-based operators will be crucial and frequent, and
in many cases the synchronization error will be completely negligible for these contacts. For all these applications, simply correcting for secular trend between TCL (or a scaled version used as Lunar time reference) and UTC could  be sufficient to time-tag Moon-related events in UTC at the microsecond level. 

For higher accuracy applications like the navigation systems, the accurate difference between TCL and UTC will be computed in the frame of time transfer between clocks on the Moon and clocks on the Earth, and will depend on the clock positions.

\section{Options for a reference Lunar Time}

Since the origin, the measurement of time on the Earth was based on the Earth rotation. The second was defined first as a fraction of the solar day. The clocks were  designed to tick this
time on the surface of the Earth. When special and general relativity effects could not be neglected anymore, with the need to define a coordinate time associated with the GCRS, it was of course natural to continue with the time rate on the Earth surface as it was in use since millennia. For this reason, the IAU defined the Terrestrial Time TT as a scaled version of TCG, so that the rate of TT is as close as possible to the rate of a clock on the surface of the Earth.
For the moon, the situation is completely different as there is currently no clock and no timescale in use there yet. Up to now, all missions to the Moon used UTC as practical time scale. Furthermore, as mentioned previously, UTC will most probably   remain the reference there in the future,  especially for the communication with the Earth. It is only for applications requiring a higher precision, as e.g. the PNT, where the nanosecond level is required, that a reference timescale has to be defined, associated with the LCRS. 

TCL could be used of course --let's call this option \eqref{item:1}. But for several practical reasons, 
also a scaled version of TCL could be considered. Let's call it Lunar Time, with the acronym TL. Two such possibilities have been studied so far.  The first one, called option \eqref{item:2} in what follows, is a scaling of TCL such that the reference time scale is close  to the rate of a clock on the surface of the Moon. It mimics how the Terrestrial Time has been defined on Earth, and requires to define a conventional lunar equipotential value like $W_0$ was set for the Earth to define TT. As seen in section \ref{ProperMoon}, the scaling factor in this case would be about $3.14 \times 10^{-11}$. Again, note that contrarily to the Earth, there is currently no timekeeping devices in use on the Moon, that would require such a definition as was the case for the TT definition. However it can be interesting to investigate the possible technical opportunities of this TL definition for future actors of Lunar exploration requiring high precision timing.

The second possible scaling considered, called option \eqref{item:3} in what follows, is such that the difference between the lunar reference time scale and UTC has no secular drift, but only periodic terms. This approach would be in the same spirit as the IAU definition of TDB which has no secular drift with respect to TT, but only periodic terms. As seen in section \ref{differences} the scaling factor corresponding to this option \eqref{item:3} would be about $6.8 \times 10^{-10}$.

These 3 options can be formalized in this general formula : 
\begin{equation}
    \label{TL}
    \TL=\TCL + \Delta f \, (\TCL-\TCL_0) + \mathrm{const.}
\end{equation}
with the value of $\Delta f$ for each option:
\begin{enumerate}
    \item $\Delta f = 0$ which means that we use directly TCL; \label{item:1}
    \item $\Delta f$ is such that the rate of TL corresponds, on average, to the rate of proper time of a clock on a given lunar geoid; in this case, $\Delta f \sim 3 \times 10^{-11}$. \label{item:2}
    \item $\Delta f$ is such that there are only periodic variations between TL and TT; in this case, $\Delta f \sim 6.8 \times 10^{-10}$.  \label{item:3}
\end{enumerate}

In our trade-off analysis, we focus  on the scaling aspects as well as on the need for clock steering in the different options. 

\subsection{Scaling}
Options \eqref{item:2} and \eqref{item:3} would require an associated scaling of mass parameters and distances, which would unavoidably introduce complexity in every computation, especially while computing Earth-Moon time and frequency transfer. In addition, because we consider that a similar approach should be used later for Mars and maybe for other planets, this would again ask for additional scalings, with, for each of them, a different numerical value for mass parameters and distances in the solar system. As mentioned before, the IAU already defined two scaled coordinate times: TT and TDB. This is associated with scaling of the mass parameters so that for example Geocentric gravitational constant $Gm_\textrm{E}$ is provided as 3 different numerical values by the IAU WG on Numerical Standards for Fundamental Astronomy \cite{IAUCBE}: TCB-compatible, TDB-compatible, and TT-compatible. Introducing a new scaling as in options \eqref{item:2} and \eqref{item:3} would bring again additional values for the same quantities and lead to possible confusion. Furthermore, the scaling of coordinate time is also associated with the corresponding scaling of distances. Options \eqref{item:2}  and \eqref{item:3} would have an effect of respectively about 1 cm  and about 30 cm on the Earth-Moon distance.
As detailed in the IERS Conventions \citep{2010ITN....36....1P},  Lunar Laser Ranging makes use of  
geocentric space coordinates for the Earth station TT-compatible (position vector $\xg^*$) that are then transformed into  
barycentric space coordinates  TDB-compatible (position vector $\xb^{**}$ ). The
transformation requires both the scaling by (1-$L_G$) followed by a rescaling by (1-$L_B$). Once LLR or any other time transfer technique will furthermore include clock reading on the Moon, in a scaled version of TCG, it will require an additional third scaling by (1-$\Delta f$) with the $\Delta f$ value given in the previous section.  We therefore state that in order to avoid  possible confusions caused by an additional scaling, option  \eqref{item:1} should be preferred.

\subsection{Steering}
The Lunar Time TL in option \eqref{item:2} is built so as to get the coordinate time having the same  rate as the proper time of a clock on the lunar surface. As seen in Figure \ref{fig:lunar_total}, the clock rate can still deviate from TL by up to $2 \times 10^{-13}$ depending on the clock  altitude with respect to the equipotential used to define TL.
Using option \eqref{item:1}, i.e. TCL as reference time, means that a perfect clock placed on the surface of the Moon would have a rate difference of  $3.14 \times 10^{-11}$ (i.e. about 2.7 $\microsec / \mathrm{day}$) with respect to the reference,  as seen in section \ref{ProperMoon}. 
Finally, for option \eqref{item:3}, the proper time of a perfect clock located on the Moon surface would have a  rate difference of about $6.5 \times 10^{-10}$ (i.e., $56\microsec / \mathrm{day}$) with respect to the reference TL. 

However, all these considerations correspond to a perfect clock. The relative frequency offsets (or rate differences) mentioned here above will only be  visible if the clock accuracy is at least of the same order of magnitude as the rate difference.
When the user wants to have its clock aligned on the reference, it is always possible to apply a steering, i.e. a frequency offset correction to the clock signal to modify its rate. This is currently done on the Earth for the majority of clocks. 

Indeed, as mentioned before, the international reference time scale UTC is the realization of TT (modulo the leap seconds). Its time unit is the SI second realized  at the reference gravity potential value $W_0$. As UTC is computed a posteriori from an ensemble of clocks and time transfer measurements, each time laboratory maintains a physical realization of UTC named UTC(k), where k is the acronym of the laboratory.  These UTC(k)s are  built from the frequency of an atomic clock which is regularly steered towards UTC to compensate both the clock error and the  relativistic redshift due to the clock  altitude above $W_0$. The GNSS satellite clocks are also steered on UTC by applying a frequency correction of about $4 \times 10^{-10}$ which compensates for  the mean relativistic effect due to the  altitude and speed of the satellite, and further regular frequency corrections are applied to compensate for the clock instability. Other user clocks can be steered using different time and frequency transfer techniques like radio signals, internet protocols, etc. Finally, we can say that the clocks are not steered on UTC only if the user need is not higher than the accuracy of its clock. We  expect a similar situation on the Moon, where the clocks  could be adjusted to the reference time broadcast for instance by the lunar navigation systems, or some ground station --on Earth in the first stage, or on the Moon in the future. However, depending on the option chosen for the reference, the magnitude of the steering could be different. The magnitude of the steering for satellite clocks will of course depend on their altitude and speed, more than on the reference time definition. For what concerns the clocks on the surface, a steering of about $3 \times 10^{-11}$ would be required for option \eqref{item:1}, around $6.5 \times 10^{-10}$ for option \eqref{item:3}, and  not more than $5 \times 10^{-13}$ for option \eqref{item:2}. However, as mentioned previously, the need for steering is determined not only by the definition of the reference but also by the clock accuracy and the user needs. Table \ref{tab:steering} indicates when the steering is needed, and what would be the magnitude of the frequency correction. The  thresholds were chosen as follows: $10^{-9}$ corresponding roughly to the scaling of option \eqref{item:3}, $10^{-11}$ corresponding roughly to the scaling of option \eqref{item:2}. From this table, we can  observe a difference between the three options only in two cases: (a) steering would be needed only for option \eqref{item:3} when the clock accuracy and the user need are both higher than the $10^{-9}$ level, and (b) option \eqref{item:2} only reduces the  magnitude of the steering correction if the clock accuracy is better than $10^{-11}$ and the user requirement higher than $10^{-11}$. The bold characters corresponds to the cases where the steering in needed due to the definition of the reference time and not due to the clock frequency accuracy. We can see that using option \eqref{item:1} requires additional steering with respect to option \eqref{item:2} only for user need and clock accuracy higher than $10^{-11}$. As we expect that, as on the Earth, most of the users on the Moon will have clocks with frequency accuracy worse than $10^{-11}$ which is the performance of the best Rubidium clocks, we can consider that only option \eqref{item:3} could represent a disadvantage concerning the steering, while both options \eqref{item:1} and \eqref{item:2} imply the same need for what concerns the steering. 

\begin{table}[t]
    \centering
    \caption{Magnitude of the steering correction, as a function of the user need}
    \vspace{0.2cm}
    \begin{tabular}{c | c c c | c c c | c c c  }
    \midrule
      clock accuracy    & \multicolumn{9}{c}{  user need (in $10^{-X}$)}\\
    (in $10^{-Y}$)  &  \multicolumn{3}{c}{$X<9$} &  \multicolumn{3}{c}{ $9 <X< 11$} &  \multicolumn{3}{c}{ $X>11$ }\\
     \midrule
    \multicolumn{1}{r}{ \textit{option}} & (1) & (2) & (3) &  (1) & (2) & (3) & (1) & (2) & (3)\\
     \midrule
     $Y<9$     & ---&---&---& <9 & <9 & <9 & <9  & <9   & <9 \\
     $9<Y< 11$ & ---&---&---& ---&--- &  \textbf{9} &9-11 & 9-11 & \textbf{9}\\
     $Y>11$    & ---&---&---& ---&--- &  \textbf{9} &  \textbf{11} &  >11 & \textbf{9}\\    
       \midrule
    \end{tabular}
\label{tab:steering}
\end{table}

\section{Conclusion}

Three options have been proposed for a practical reference time on the Moon:  \eqref{item:1} using  TCL, or
using a scaled version of TCL so that  \eqref{item:2} the lunar reference timescale corresponds, on
average, to the proper time of a clock on the surface of the Moon, or,  \eqref{item:3} the lunar
reference timescale differs from TT only by periodic variations. We have reviewed these 3 options 
under the considerations of scaling and steering. 

We have shown that the scaling of TCL associated with  options  \eqref{item:2} and  \eqref{item:3} would require an associated scaling of mass parameters
and distances, which would unavoidably introduce complexity in every computation,
especially while computing Earth-Moon time and frequency transfers.   

For what concerns the steering, even if  option \eqref{item:2} if envisaged as it would allow  that an accurate clock on
the Moon surface would tick the reference time TL, this argument is valid only
at the condition that the clock accuracy is better than $10^{-11}$. 
For less accurate clocks or more precise user needs, the clocks should be
steered on the reference, exactly as if the reference is not defined on a lunar geoid. Furthermore, all users on the Moon who need to communicate with the Earth will probably steer their clock on average on UTC, i.e. a steering of about $6.5 \times 10^{-10}$. In
view of this, and considering in addition that option  \eqref{item:2} would require the definition
of a lunar $W_0$ which does not exist to date, we consider that the less constraining and
natural solution for the Lunar Reference Time  is to use option  \eqref{item:1}, that is to say using TCL.

\bibliographystyle{plainnat}
\bibliography{lunar_time}

\end{document}